\documentclass[aps,prb,twocolumn,superscriptaddress,showpacs,amsmath]{revtex4}
\usepackage{graphicx}

\begin{document}

\title{$0$-$\pi$ transition in SFS junctions with strongly spin-dependent
scattering}
\author{O. Kashuba}
\affiliation{Department of Physics, Lancaster University, Lancaster, LA1 4YB, UK}
\author{Ya. M. Blanter}
\affiliation{Kavli Institute of NanoScience, Delft University of Technology, Lorentzweg 1, 2628 CJ Delft, The Netherlands}
\author{Vladimir I. Fal'ko}
\affiliation{Department of Physics, Lancaster University, Lancaster, LA1 4YB, UK}

\begin{abstract}
We develop theory of the critical current in a superconductor -- ferromagnetic alloy --
superconductor trilayers, which takes into account strong spin dependence of
electron scattering of compositional disorder in a diluted ferromagnetic
alloy. We show that in such a system the critical current oscillations as
the function of the thickness of the ferromagnetic layer, with the period of 
$v_{F}/2I$, $v_F$ and $I$ being the Fermi velocity and exchange splitting, respectively,
decay exponentially with the characteristic length of the order of the mean free path.
\end{abstract}

\maketitle


The recent observation~\cite{ryazanov:T,kontos:dos,blum:df,Blamire} of
Josephson junctions with negative coupling,~\cite{bulaevskii:sf,Golubov}
also known as $\pi$ junctions, has attracted a lot of attention to hybrid
superconductor -- ferromagnet -- superconductor (SFS) structures. In
contrast to conventional Josephson junctions, such as superconductor --
normal metal systems, where the ground state corresponds to the
superconducting phase difference $\varphi$ of zero, the phase difference
in a SFS trilayer can take both $\varphi=0$ and $\pi$ values,
depending on the thickness of the ferromagnetic layer. Both $0$ and $\pi$
states in SFS trilayers have been deduced from the measurements of the
density of states~\cite{kontos:dos} and the critical current as a function
of magnetic flux and temperature.~\cite{kontos:phase,ryazanov:T,sellier:T,Blamire}
In particular, the critical current exhibits oscillations superimposed on
the exponential decay as a function of the thickness of the ferromagnetic
layer.~\cite{kontos:df,blum:df}
The decay length $\xi _{d}$ and the period $\xi _{o}$ of these
oscillations have been measured, providing comparable yet unequal
experimental values of these two parameters.

In ballistic SFS structures, the critical current is expected to oscillate
with the period $\xi _{o}=v_{F}/2I$. For a ferromagnetic metal with a strong
exchange splitting $I$, fluctuations of the width of the ferromagnetic layer
suppress the appearance of the proximity effect, despite the fact that in
ballistic structures Cooper pairs decay with the distance according to a
power law rather than exponentially. Moreover, it has been shown~\cite{volkov}
that, when the electron motion in a ferromagnetic film with large $I$
is diffusive, the randomisation of the oscillation phase over paths of
different lengths leads to the expenential suppression of proximity at the
length of the mean free path: $\xi_{d}\sim l$ for the case of $I\tau\gg 1$
(where $\tau$ is the electron mean free path). To enhance the proximity
effect in a SFS multilayers, one may want to use weakly ferromagnetic
alloys, where the exchange field $I$ is reduced by diluting the magnetic
component. The analysis of diluted systems with $I\tau \ll 1$ based upon
modelling disorder in SFS junctions as spin-independent impurities has shown
that the decay length may be expected~\cite{buzdin:sf,Golubov} to extend
beyond the mean free path range, such that $\xi_{d}\sim \xi_{o}=\sqrt{D/I}$,
where $D=v_{F}^{2}\tau /3$.

In this paper, we show that a possibility to prolong the extent of the
superconducting proximity effect in SFS structures by making them of diluted
magnetic alloys is strongly limited. Following theory of suppression of
superconductivity by magnetic impurities~\cite{Abrikosov}, earlier
theories~\cite{BuzdinPR,Golubov1} took into account the effect of magnetic disorder
by including in the Usadel equation a weak Cooper pair relaxation described
by a phenomenological spin relaxation rate $\tau_{s}^{-1}$. Keeping in mind
that even in a weak ferromagnet electron spin flip is an inelastic process
and should be accompanied by the excitation of a magnon, we attribute the
pair breaking in a ferromagnetic alloy to a giant magnetoresistance (GMR) 
type effect. As noticed in earlier GMR studies~\cite{Zhang,Fert}, a feature of
ferromagnetic alloys is that elastic electron scattering in them is strongly spin
dependent.
Indeed, one scattering event off strongly spin-dependent disorder, seen differently
by spin-up and spin-down electrons, is enough to break a singlet Cooper
pair. In such a case, the decay length of a Cooper pair is of the order of
the mean free path, $\xi_{d}\sim l$. Since, in this case, the use of Usadel
equations adopted in the previous studies of disordered SFS
junctions~\cite{buzdin:sf,Golubov} does not hold, here we employ a nonlocal
approach based on solution of Eilenberger 
equation~\cite{Eilenb,Golubov,BuzdinPR,volkov} to describe $0$-$\pi $ Josephson
oscillations as a function of the thickness of the diluted ferromagnetic
alloy layer.

To describe a dilute ferromagnetic alloy, we use the following Hamiltonian
(a $2 \times 2$ matrix in the spin space), adopted~\cite{Zhang} in GMR
theory, 
\begin{equation}  \label{eq:impurity}
\mathcal{H} = \hat p^2/2m + V(\mathbf{r}) +\boldsymbol{\sigma} \mathbf{J}
(\mathbf{r}) \ ,
\end{equation}
where $V$ and $\mathbf{J}$ describe magnetic atoms embedded into a
normal metal, and $\boldsymbol{\sigma}$ is the vector of Pauli matrices. The
average $\langle \mathbf{J} \rangle = \mathbf{e}_zI$ determines the
exchange splitting for conduction band electrons, and $\langle V \rangle = 0$.
Since every magnetic atom produces both scalar $V$ and exchange $\mathbf{J}$
potentials, we use the following correlation functions for
magnetic and nonmagnetic disorder,
$\langle V (\mathbf{r}) V (\mathbf{r'}) \rangle =
(2\pi\nu\tau_V)^{-1} \delta(\mathbf{r}-\mathbf{r}')$,
$\langle J_{\alpha}(\mathbf{r}) J_{\beta}(\mathbf{r}') \rangle =
(2\pi\nu\tau_J)^{-1} \delta_{\alpha\beta} \delta(\mathbf{r}-\mathbf{r}')$,
and $\langle V(\mathbf{r}) J_{\alpha}(\mathbf{r}') \rangle =
(2\pi\nu\tau_\mathsf{mix})^{-1} \delta_{\alpha z}\delta(\mathbf{r}-\mathbf{r}')$.


The starting point for quantitative description is Eilenberger equation for
the retarded component of the semi-classical Green's function, 
\begin{gather}
v_F\mathbf{n}\partial_\mathbf{r}\check{g} + \left[-i\omega\tau^3 + iI\sigma^z
+ i\check{\Sigma},\check{g}\right]_-=0, \label{eq:eilenberger}\\
\check{g} \equiv
\begin{pmatrix}
g & f \\
f^+ & -g
\end{pmatrix}, \,\, f \equiv
\begin{pmatrix}
0 & f_{\uparrow\downarrow} \\
f_{\downarrow\uparrow} & 0
\end{pmatrix}, \,\, \check{g}^2=1,
\end{gather}
where $(f^+)_{\alpha\beta} (\mathbf{r}, t; \mathbf{n}, \omega) =
-[f_{\alpha\beta} (\mathbf{r}, t; -\mathbf{n}, -\omega)]^*$, $\tau^3$ acts
in the Nambu space, $\mathbf{n} = \mathbf{p}/p$, the self-energy has the
form
\begin{equation*}
i\check{\Sigma}=\frac{1}{2\tau_V}\langle\check{g}\rangle + \frac{1}{2\tau_%
\mathsf{mix}}\left[\sigma^z,\langle\check{g}\rangle\right]_+ + \frac{1}{%
2\tau_J}\sigma^z\langle\check{g}\rangle\sigma^z,
\end{equation*}
and $\langle\check{g}\rangle=\int\check{g}\,d^2\mathbf{n}/4\pi$ is the
Green's function averaged over momentum direction. For the weak proximity
effect, Eq.~(\ref{eq:eilenberger}) can be linearized around the zero-order
Green's function $\check{g}_0=\tau^3$. Performing the expansion up to first
order, we obtain
\begin{multline}
v_F\mathbf{n}\partial_\mathbf{r} f - 2i\omega f + 2iI \sigma_z f + \\
\left(\tau_V^{-1}+\tau_J^{-1}\right) f -
\left(\tau_V^{-1}-\tau_J^{-1}\right) \langle f\rangle = 0.
\label{eq:kin_f}
\end{multline}

The linearization of the Eilenberger equation and subsequent analysis are based
upon the assumption of weak coupling between superconductors and the
ferromagnet, which is realized, for instance, if these are separated by an
opaque barrier with the low transparency $\Theta \ll 1$. The appropriate
boundary conditions have been derived by Zaitsev,~\cite{zaitsev}
\begin{align*}
&\check{g}_{S}^{a}=\check{g}_{F}^{a}\equiv \check{g}^{a},\\
&\check{g}^{a}\{(1-\Theta )(1-(\check{g}^{a})^{2}]+\Theta (\check{g}_{-}^{s})^{2}\}=
\Theta \check{g}_{-}^{s}\check{g}_{+}^{s},
\end{align*}
where $\check{g}_{i}^{s/a}=(\check{g}_{i}(n_{z})\pm \check{g}_{i}(-n_{z}))/2$
($i=S$ and $i=F$ for a superconductor and a ferromagnetic alloy,
respectively), $n_{z}>0$, where $n_{z}$ is the projection of
$\mathbf{n}=\mathbf{p}/p$ onto the direction normal to the SF interface,
and $\check{g}_{\pm }^{s}=(\check{g}_{S}^{s}\pm \check{g}_{F}^{s})/2$.
In the case of low transparency $\Theta \ll 1$, we find that in the first
order in $\Theta $,
\begin{equation*}
\check{g}^{a}=\frac{\Theta }{4}[\check{g}_{S}^{(0)},\check{g}_{F}^{(0)}]_{-},
\end{equation*}
where $\check{g}_{S}^{(0)}$ and $\check{g}_{F}^{(0)}$ are the Green's
functions in the two materials when those are detached ($\Theta=0$).
Together with Eq.~(\ref{eq:kin_f}) this gives us closed set of equations.


It is convenient to represent the semiclassical Green's function
$f(\mathbf{n},z)$ as a combination of two functions of a positive argument,
$n_{z}>0$: $f_{1}(n_{z},z)\equiv f(n_{z},z)$ and $f_{2}(n_{z},z)\equiv f(-n_{z},z)$.
In this representation the boundary conditions take the form
\begin{gather}
f_{1}(n_{z},0)-f_{2}(n_{z},0) =a_{L}, \nonumber \\
f_{1}(n_{z},d_{F})-f_{2}(n_{z},d_{F}) =-a_{R}, \nonumber \\
a_{L/R} =-\Theta \Delta \exp (i\phi _{L/R})/\sqrt{\Delta ^{2}-\omega ^{2}},
\end{gather}
where $d_{F}$ is the thickness of the ferromagnetic layer. The equations for 
$f_{1}$ and $f_{2}$ take the form 
\begin{equation}
\begin{split}
& n_{z}\partial_{z}f_{1}(n_{z},z)+\lambda f_{1}(n_{z},z)-\alpha \langle
f(z)\rangle =0, \\
-& n_{z}\partial_{z}f_{2}(n_{z},z)+\lambda f_{2}(n_{z},z)-\alpha \langle
f(z)\rangle =0,
\end{split}
\label{eq:mathset}
\end{equation}%
where $n_{z}>0$ and $\alpha =(\tau _{J}-\tau _{V})/(\tau _{V}+\tau _{J})$,
$-1\leq \alpha \leq 1$. Also,
\begin{equation}
\lambda =1-2i(\omega - I\sigma_{z})\tau ,
\end{equation}
is a $2\times 2$ matrix acting on the $2\times 2$ matrix $f$, and $\tau
^{-1}=\tau _{V}^{-1}+\tau _{J}^{-1}$. The averaged Green's function equals 
\begin{equation}
\langle f(z)\rangle =\int_{0}^{1}dn_{z}\,\bigr(f_{1}(n_{z},z)+f_{2}(n_{z},z)%
\bigr).
\end{equation}

In the case of a thick ferromagnetic layer, such that $e^{-d_F/l}\ll 1$,
where $l = v_F\tau$ is the mean free path, one can write down the formal
solution of Eqs.~(\ref{eq:mathset}) as
\begin{subequations}
\label{eq:f_solution}
\begin{multline}
f_1(n_z,z)=a_L e^{-\lambda z/n_zl} + \frac{\alpha}{l} \int_{0}^{z}
e^{\lambda (z^{\prime}-z)/n_zl} \langle f (z^{\prime}) \rangle \frac{%
dz^{\prime}}{n_z} \\
+ \frac{\alpha}{l} \int_{0}^{d_F} e^{-\lambda (z+z^{\prime})/n_zl} \langle f
(z^{\prime})\rangle \frac{dz^{\prime}}{n_z} ,
\end{multline}
\begin{multline}
f_2(n_z,z)=a_R e^{\lambda (z-d_F)/n_zl} \\
+ \frac{\alpha}{l} \int_{z}^{d_F} e^{\lambda (z-z^{\prime})/n_zl} \langle f
(z^{\prime}) \rangle \frac{dz^{\prime}}{n_z} \\
+\frac{\alpha}{l} \int_{0}^{d_F} e^{\lambda (z^{\prime}+z-2d_F)/n_zl}
\langle f (z^{\prime}) \rangle \frac{dz^{\prime}}{n_z}.
\end{multline}

The subsequent algebra includes adding and averaging Eqs.~(\ref{eq:f_solution}),
which leads to the integral equation for $\langle f\rangle$.
Having presented $\langle f(z)\rangle $ as the sum,
\end{subequations}
\begin{equation}
\langle f(z)\rangle =a_{L}h(z)+a_{R}h(d_{F}-z)\ ,  \label{eqh}
\end{equation}
we find that the (matrix) function $h(z)$ satisfies Fredholm equation of the
second type,
\begin{multline}
2h(z)=K(\lambda z)+(\alpha /l)\int_{0}^{d_{F}}\Bigl(G\bigl(\lambda
|z-z^{\prime }|\bigr)+  \label{eq:int_eqn} \\
G\bigl(\lambda (z+z^{\prime })\bigr)+G\bigl(\lambda (2d_{F}-z-z^{\prime })%
\bigr)\Bigr)h(z^{\prime })dz^{\prime },
\end{multline}%
where $G(z)=\int_{0}^{1}n_{z}^{-1}e^{-z/n_{z}l}dn_{z}$; $K(z)=%
\int_{0}^{1}e^{-z/n_{z}l}dn_{z}$.

Up to this point, we could still reduce our equations to Usadel equations provided the diffusion approximation holds, $(1-\alpha )||\mathop{\mathrm{Im}}\nolimits \lambda ||\ll 1$.
In the rest of the paper, we work outside this regime and consider the ballistic situation.
For $\alpha =0$, the exact solution of Eq. (\ref{eq:int_eqn}) is $h(z)=K(\lambda z)/2$.
Generalizing, we find that in the ballistic case the solution is determined by behavior of functions $K$ and $G$ which at $z\gg l$ are $K(z)\approx G(z)\approx e^{-z/l}l/z$.
Assuming that solution falls off exponentially as $e^{-\lambda z/l}$, one can see that in Eq.~(\ref{eq:int_eqn}) the last term in the integral can be neglected everywhere except for a small region near the boundary, $z=d_{F}$.
This enables us to split the solution of Eq.~(\ref{eq:int_eqn}) into two parts,
\begin{equation}
h(z)=e^{-\lambda z/l}h_{L}(z)+e^{-\lambda (2d_{F}-z)/l}h_{R}(d_{F}-z).
\end{equation}

The first term is relevant everywhere and is the main term of the solution,
whereas the second one is only important close to the boundary, $d_{F}-z\sim
l$, when the exponents become of the same order. Each of the matrix
functions $h_{i}$ ($i=L,R$) satisfies the equation
\begin{widetext}
\begin{equation}
2h_{i}(z)=S_{i}(\lambda z)+\frac{\alpha}{l}\int_{0}^{z}\tilde{G}(\lambda(z-z'))h_{i}(z^{\prime })dz'+
\frac{\alpha}{l}\int_{0}^{\infty}e^{-2\lambda z'}
\Bigl(\tilde{G}(\lambda (z+z'))h_{i}(z^{\prime }) + \tilde{G}(\lambda z')h_{i}(z+z')\Bigr)dz',
\label{eq:eqn_h}
\end{equation}
\end{widetext}

where $\tilde{G}(\lambda z)=G(\lambda z)e^{\lambda z/l}$ and $S_{L}(\lambda
z)=K(\lambda z)\exp (\lambda z)$.

Far from the left boundary, $z\gg l$, we parameterize $h_{L}(z)=A(z)l/\lambda z$,
$1/\lambda\equiv\lambda^{-1}$.
Substituting it into Eq.~(\ref{eq:eqn_h}) and keeping the
leading order in $l/z$, we obtain the equation for the diagonal matrix $A$,
\begin{multline}
\left( 2-\frac{\alpha }{\lambda }\ln 2\right) A(z)=
1+\frac{\alpha}{\lambda }\xi
+\frac{\alpha }{\lambda }\int_{0}^{z}\frac{A(z^{\prime })}{z'}dz' \\
+\frac{\alpha }{\lambda l}\int_{0}^{z}A(z-z')\tilde{G}(\lambda z')dz',
\label{eqn_h1}
\end{multline}
where the matrix $\xi =\lambda \int_{0}^{\infty }h_{L}(z)e^{-2\lambda z}dz/l$
does not depend on $z$. The last term in Eq.~(\ref{eqn_h1}) in the leading
order in $\ln ^{-1}z$ is $A(z)(\gamma +\ln (\lambda z/l))$ with $\gamma $
being Euler's constant. Subsequently, we obtain a differential equation for
the function $\int_{0}^{z}A(z')dz'/z'$. The solution far from the boundaries reads 
\begin{equation}
\begin{split}
&h_{L}(z)=A(z)\frac{le^{-\lambda z/l}}{\lambda z}, \\
&A(z)=\frac{\delta(\alpha ,\lambda )}{\bigl[2-(\alpha /\lambda)(\gamma +\log 2\lambda z/l)\bigr]^{2}},
\end{split}
\end{equation}
where a constant $\delta (\alpha ,\lambda )$ is of order one; at $\alpha =0$
the exact solution gives $\delta (0,\lambda )=2$. Numerical calculations
show that $\delta (\alpha ,\lambda )$ is still close to $2$ even for 
$\alpha=1$.

Having solved the equation for $h_{L}$, we use it to determine the matrix
function $S_{R}$, according to 
\begin{eqnarray*}
S_{R}(z) &=&\frac{\alpha }{l}\int_{0}^{d_{F}}\tilde{G}\left( \lambda
(z+z^{\prime })\right) \\
&\times &\left[ h_{L}(d_{F}-z^{\prime })-e^{-2\lambda z^{\prime
}/l}h_{L}(d_{F}+z^{\prime })\right] dz^{\prime }\ ,
\end{eqnarray*}%
and find the solution for the function $h_{R}(z)$.

\begin{figure}[tbp]
\centering
\includegraphics[scale=0.7]{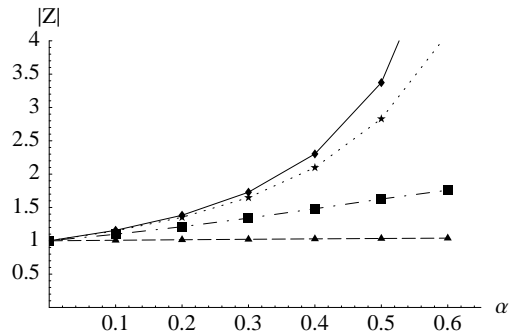}
\caption{\small Dependence of the factor $|Z|$ determined in Eq.~(\ref{Z}) on the $\alpha = (\tau_J-\tau_V)/(\tau_V+\tau_J)$ for various values of $I\tau$.
From top to bottom: $I\tau=0,0.3,1,10$.
The results show that outside of the diffusive regime, $I\tau\ll 1$ and $\tau_J\gg\tau_V$, $Z$ is a smooth function of $\alpha$ of order of 1.}
\label{fig:alpha}
\end{figure}

Within the approximations used in the above analysis of the Eilenberger
equation for the anomalous Green function $f$, the Josephson current density
in the SFS structure can be represented as 
\begin{equation}
\mathbf{j}=-\pi e\nu v_{F}\int n(\omega )\mathop{\mathrm{Re}}\nolimits \mathcal{I}\,
\frac{d\omega }{2\pi },\quad \mathcal{I}=\left\langle \mathbf{n}\,
\mbox{\rm tr}\bigl(\hat{f}\hat{f}^{+}\bigr)\right\rangle.
\end{equation}
Here, $n(\omega )$ is the Fermi distribution function. Substituting the
expressions for $h(z)$ into Eq. (\ref{eqh}) and Eqs. (\ref{eq:f_solution}),
we find 
\begin{equation}
\mathcal{I}=\frac{\Theta ^{2}\Delta ^{2}\sin (\phi _{L}-\phi _{R})}{\Delta
^{2}-\omega ^{2}}\mbox{\rm tr}\frac{le^{-\lambda d_{F}/l}}{\lambda d_{F}}%
Z(\alpha ,\lambda ,d_{F}),  \label{eq:current}
\end{equation}%
where the matrix function $Z$ depends on $d_{F}$ logarithmically and for $%
\alpha =0$ equals $Z(0,\lambda ,d_{F})=1$. Generally, in the leading order
in $d_{F}^{-1}$ it becomes 
\begin{multline}
Z=\frac{\alpha }{\lambda }\xi \left( 1+\frac{\alpha }{\lambda }\xi \right) +
\frac{\alpha ^{2}}{\lambda }\zeta + \\
\left( 2-\frac{\alpha }{\lambda }\left( 1+\ln 2\right) \right) A(d_{F}),
\label{Z}
\end{multline}
where $\xi =\lambda \int_{0}^{\infty }h_{L}(z)e^{-2\lambda z}dz/l$ and
$\zeta =\alpha ^{-1}\lambda ^{2}d_{F}\int_{0}^{\infty }h_{R}(z)K(\lambda z)dz/l$.
For $d_{F}\gg l$ the quantity $\xi $ is constant, and $A$ depends
on it logarithmically. The quantity $\zeta $ depends on $d_{F}$ in the same
way as $A$. For $\Delta \tau \ll ||\lambda ||\sim 1$, the calculation of the
frequency integral leads to an expression that is most conveniently
represented as the sum over Matsubara frequencies $\omega _{n}=2\pi T(n+1/2)$.
This is equivalent to the replacement $i\omega \rightarrow \omega_{n}$ in
the above expressions involving the matrix $\lambda $. As the result,
$\lambda $ becomes a diagonal matrix with two complex conjugate eigenvalues,
so that in the Matsubara representation $Z$ has the same property and can be
written down as
\begin{equation*}
Z=|Z|\exp (i\sigma_{z}\varphi_{Z})\ .
\end{equation*}
Note that for $\omega_{n}\tau \lesssim \Delta \tau \lesssim 1$, $|Z|$ and
$\varphi_{Z}$ are two parameters of a structure independent of the Matsubara
frequency. For $\alpha =0$, one finds $Z=1$. The dependence of $|Z|$ on the
value of parameter $\alpha $ is plotted in Fig.~\ref{fig:alpha}.
Although parameters $|Z|$ and $\varphi_{Z}$ depend on the quantities
$\alpha =(\tau_{J}-\tau_{V})/(\tau _{V}+\tau _{J})$, $I\tau $, and $d_{F}/l$,
this fact does not qualitatively affect the results. Outside the regimes of
$I\tau \ll 1$ and $\tau_{J}\ll \tau _{V}$ (where our results are not applicable) $Z$
is a smooth function of $d_{F}$ of the order 1 that does not contain any
dependence on scales of order of $\xi_{d}\sim l$ or $\xi_{o}=v_{F}/2I$.

Finally, we arrive at the expression for the critical current density [in
$j=j_{c}\sin (\phi_{L}-\phi_{R})$] which has the form
\begin{multline}
j_{c}=2\pi e\nu v_{F}\Theta ^{2}|Z|\frac{e^{-d_{F}/l}}{d_{F}/l}T\sum_{\omega
_{n}>0}\frac{e^{-2\omega _{n}d_{F}/v_{F}}}{1+\omega _{n}^{2}/\Delta ^{2}}%
\times   \label{eq:result} \\
\frac{\cos \left( 2Id_{F}/v_{F}+\arctan (2I\tau )-\varphi _{Z}\right) }{%
\sqrt{1+(2I\tau )^{2}}}.
\end{multline}
In the limiting cases, the summation over Matsubara frequencies $\omega_{n}$
can be calculated explicitly. For a ferromagnetic layer with the thickness
much greater than the coherence length in the superconductor,
$d_{F}\gg v_{F}/\Delta $, the sum equals $2T\sinh ^{-1}(2\pi Td_{F}/v_{F})$.
In the opposite case of a thin layer, $d_{F}\ll v_{F}/\Delta $, one obtains
$(\Delta/4)\tanh(\Delta/2T))$. At zero temperature, the sum can be
converted into integral which equals $\mathop{\mathrm{Ci}}\nolimits(a)\sin(a)
+(\pi/2-\mathop{\mathrm{Si}}\nolimits(a))\cos (a)$, where
$a=2\Delta d_{F}/v_{F}$, and the functions $\mathop{\mathrm{Si}}\nolimits$ and
$\mathop{\mathrm{Ci}}\nolimits$ are sine and cosine integrals, respectively.
For high temperature, $d_{F}\gg v_{F}/T$, only the lowest frequency is
important, and the sum equals $T(1+\pi^2T^2/\Delta^2)^{-1}\exp(-2\pi T d_{F}/v_{F})$.

\begin{figure}[tbp]
\centering
\includegraphics[scale=0.7]{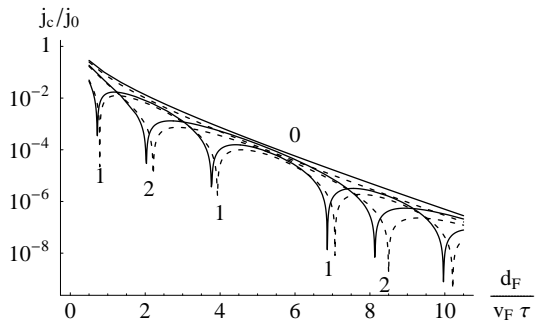}
\caption{{\protect\small The dependence of the critical current in SFS trilayer on ferromagnetic layer thickness.
Here, we normalize current using a notional $j_0 = 2\pi e\nu v_F \Theta^2 \Delta$, and show
it for $T=0$, $\Delta \tau =0.1$ and several values of parameters $\alpha$ and $I\tau$: $\alpha=0$ (dashed lines) and $\alpha=0.3$ (solid lines).
Dips in the value of $j_c$ indicate positions where it disappears and changes sign, thus resulting in a sequence of $0-\pi$ transitions.
Neighbouring dips always correspond to the same value of $I\tau$, demonstrating only weak dependence of the results on the parameter $\alpha$.
For comparison, we also show the decay of the Josephson proximity effect in an SNS structure heavily doped by magnetic scatterers ($I\tau=0$).}}
\label{fig:currents}
\end{figure}

The critical current dependence on the ferromagnetic layer thickness
described in Eq.~(\ref{eq:result}) for a weakly ferromagnetic layer with
$d_{F}>l$ is shown in Fig.~\ref{fig:currents}.
Even a when dilution of a ferromagnetic layer is such that exchange energy in
it is weak, $I\tau \ll 1$, oscillations of $j_{c}$ as a function of the layer
thickness, with the period of $\xi_{o}=v_{F}/2I$, decay exponentially at the
length scale of the mean free path, $\xi_{d}=l$ -- similarly to what happens in a
disordered ferromagnetic layer with a strong exchange~\cite{volkov} ($I\tau \gg 1$).
Our results for $I\tau \gg 1$ coinscide with whose of Ref.~\onlinecite{volkov}:
For strong fields, the phase randomization of the order parameter is effective
irrespective of the nature of scatterers.
The dependence of the critical current on the thickness of the ferromagnetic
layer in Eq.~(\ref{eq:result}) resembles the experimentally observed
suppression of the proximity effect, at the length scale comparable to the
mean free path measured in the same material~\cite{ryazanov:T}. Note that
theories involving generation of the triplet order parameter due to
nonuniform (spiral) magnetization in the ferromagnet~\cite{efetov:triplet}
end up with the opposite conclusion, predicting weaker decay of the order
parameter.

In conclusion, we developed theory of the proximity effect in a superconductor
-- weakly ferromagnetic GMR alloy -- superconductor trilayers, which takes
into account strong spin dependence of electron scattering of compositional
disorder. The result, Eq.~(\ref{eq:result}), describes $0$--$\pi$
transition for Josephson effect as a function of the thickness of the
ferromagnetic layer $d_{F}$: Oscillations occur with the period of $\xi_{o}=v_{F}/2I$
and exponential decay with the characteristic length $\xi_{d}=l$ of
the order of the mean free path, even in the regime when  $I\tau\ll 1$.
This result complements previous studies of the spin-singlet
proximity effect in superconductor -- ferromagnet hybrid structures
performed for ballistic and diffusive systems with spin-independent
scattering~\cite{Golubov,BuzdinPR,volkov} as well as theories of the
suppression of the order parameter oscillations caused by spin-active
interfaces.~\cite{cottet}

The authors acknowledge useful discussion with K.Efetov, the late A. Larkin, Yu.
Nazarov, V. Ryazanov, and A. Volkov. This work was supported by EC Grant No.
NMP2-CT2003-505587 and the Lancaster-EPSRC Portfolio Partnership No. EP/C511743.

\end{document}